\documentclass[a4paper,fleqn]{cas-sc}

\usepackage[numbers]{natbib}

\def\tsc#1{\csdef{#1}{\textsc{\lowercase{#1}}\xspace}}
\tsc{WGM}
\tsc{QE}
\tsc{EP}
\tsc{PMS}
\tsc{BEC}
\tsc{DE}
\usepackage{graphicx}%
\usepackage{multirow}%
\usepackage{amssymb,amsfonts}%
\usepackage{amsthm}%
\usepackage{mathrsfs}%
\usepackage[title]{appendix}%
\usepackage{xcolor}%
\usepackage{textcomp}%
\usepackage{manyfoot}%
\usepackage{booktabs}%
\usepackage{algorithm}%
\usepackage{algorithmicx}%
\usepackage{algpseudocode}%
\usepackage{listings}%
\usepackage{hyperref}%
\usepackage{bm}

\begin{document}
\let\WriteBookmarks\relax
\def\floatpagepagefraction{1}
\def\textpagefraction{.001}
\shorttitle{Observational constraints on a modified-gravity model}
\shortauthors{M.A. Acero, A. Oliveros}

\title [mode = title]{Observational constraints on a modified-gravity model with an exponential function of the curvature using the expansion history, the RSD, and the Pantheon+SH0ES data}

\author{Mario A. Acero}[type=editor,
                        auid=000,bioid=1,
                        orcid=0000-0002-0835-0641]
\cormark[1]
\ead{marioacero@mail.uniatlantico.edu.co}
\ead[url]{https://maaceroo.github.io/}

\author{A. Oliveros}[type=editor,
                        auid=000,bioid=1,
                        orcid=0000-0002-6796-1784]

\ead{alexanderoliveros@mail.uniatlantico.edu.co}

\affiliation{organization={Programa de Física, Universidad del Atlántico},
                addressline={Carrera 30 8-49}, 
                city={Puerto Colombia},
                state={Atlántico},
                country={Colombia}}

\cortext[cor1]{Corresponding author}

\begin{abstract}
Considering a well-motivated $f(R)$ modified-gravity model, in which an exponential function of the curvature is included, in this paper we implement a statistical data analysis to set constraints on the parameters of the model, taking into account an analytic approximate solution for the expansion rate, $H(z)$. Using a Monte Carlo Markov Chain-based analysis of the expansion rate evolution, the standardized SN distance modulus and the redshift space distortion observational data, we find that the preferred value for the perturbative parameter, $b$, quantifying the deviation of the $f(R)$ model from $\Lambda$CDM, lies in a region that excludes $b = 0$ at $\gtrsim 3.5 \sigma$ C.L., and that the predicted current value of the Hubble parameter, $H_0$, locates in between the two observational results currently under scrutiny from Planck and SH0ES collaborations. Under the implemented approximate solution, and with the constraints obtained for the parameters, the proposed $f(R)$ model successfully reproduces the observational data and the predicted evolution of interesting cosmological parameters resemble the results of $\Lambda$CDM, as expected, while an oscillatory behavior of the dark energy equation of state is observed, pointing to deviation from the concordance cosmological model. The results presented here reinforce the conclusion that the $f(R)$ modified-gravity model represents a viable alternative to describe the evolution of the Universe, avoiding the challenges faced by $\Lambda$CDM.
\end{abstract}


\maketitle

\section{Introduction}\label{sec_intro}
Although Einstein's General Relativity (GR) has been enormously successful in explaining many observations at the astrophysical and cosmological levels, there are phenomena that cannot be adequately explained within this framework. For example, the current observed accelerating expansion of the Universe poses a challenge \cite{SupernovaSearchTeam:1998fmf, SupernovaCosmologyProject:1998vns}. A first attempt to explain this late-time cosmic acceleration is the introduction of a new energy component in the Universe, known as dark energy (DE), characterized by a negative pressure. However, this proposal has proven to be very difficult to incorporate within the known theories of physics (for a comprehensive review about this topic, see Refs.~\cite{Peebles:2002gy,Copeland:2006wr, Sahni:2006pa, Bamba:2012cp}). DE is commonly associated with a cosmological constant ($\Lambda$), which drives the late-time cosmic evolution and whose origins are traced back to early quantum fluctuations of the vacuum. Still, this model (known as $\Lambda$CDM) faces challenges such as the coincidence and cosmological constant problems, as well as tensions that have arisen among recent cosmological measurements. 

In order to circumvent the above issues, an interesting proposal is the $f(R)$ gravity theories, in which the Einstein-Hilbert action is modified with a general function of the Ricci scalar $R$, $f(R)$. However, the selection of a specific function $f(R)$ is not arbitrary: it must adhere to several consistency requirements and various constraints that impose conditions for the cosmological viability of $f(R)$ models. One crucial requirement is that a given $f(R)$ adequately describes the different cosmic eras, including the radiation, matter, and dark energy eras, and probably the inflationary period. Moreover, it is imperative also that the selected $f(R)$ function satisfies both the cosmological constraints and the local gravity constraints, in addition to other relevant considerations \cite{Amendola:2006we}. Numerous works have been undertaken in this context, exploring various aspects at both the astrophysical and cosmological levels \cite{Hwang:2001pu, Nojiri:2003ft, Capozziello:2005ku, Cognola:2005de, Capozziello:2004vh, Nojiri:2006gh, Song:2006ej, Mao:2006ub, Olmo:2006eh, Hu:2007tf, Faraoni:2007yn, Bean:2006up, Nojiri:2007as, Capozziello:2007ms, Sasaki:2007pt, Appleby:2008tv, Carloni:2007yv, Capozziello:2008qc, Elizalde:2010wg, Capozziello:2012ie, Odintsov:2019mlf, Odintsov:2019evb, Odintsov:2020nwm, Oikonomou:2020oex, Oikonomou:2020qah}. In general, for a more extensive review of $f(R)$ theories, interested readers are invited to see Refs.~\cite{Odintsov:2010wj,Clifton:2011jh, Odintsov:2017ncd}. 

Within this wide plethora of viable $f(R)$ gravity models, two of those have been highlighted in the literature: the Hu and Sawicki (HS) model \cite{Hu:2007nk}, and the Starobinsky model \cite{Starobinsky:2007hu}. Although these were originally advertised as models that avoid the inclusion of the cosmological constant making $f(R)$ distinct from the $\Lambda$CDM form, $f(R)=R-2\Lambda$, in Ref.~\cite{Basilakos:2013nfa} it was demonstrated that these models can be arbitrarily close to $\Lambda$CDM (where the deviation from $\Lambda$CDM is characterized by a parameter $b$), and provide predictions that are similar to those of the usual (scalar field) DE models, particularly concerning the cosmic history, including the presence of the matter era, the stability of cosmological perturbations, the stability of the late de Sitter point, etc. The authors also found that $b\sim \mathcal{O}(10^{-1})$ for the HS model, making it practically indistinguishable from $\Lambda$CDM at the background level. 

In a related investigation, in Ref.~\cite{Nesseris:2017njc}, the authors introduce a new class of models that are variants of the HS model that interpolate between the cosmological constant model and a matter-dominated universe for different values of the parameter $b$, which is usually expected to be small for viable models and which, in practice, measures the deviation from GR. Recently, in Ref.~\cite{Kumar:2023bqj}, the state-of-the-art BAO+BBN data and the most recent Type Ia supernovae (SNe Ia) sample, PantheonPlus, including the Cepheid host distances and covariance from SH0ES samples, were used to robustly constrain the HS and Starobinsky models, and found that both models are consistent with GR at the 95\% CL.

As a further contribution to the research on this matter, in this work we analyze the parameters governing the $f(R)$ model proposed in \cite{Oliveros:2023ewl} with the approximate analytical solution found by one of the authors \cite{Oliveros:2023ouq} (Section \ref{model}), setting constraints on the characteristic parameter, $b$, and on some cosmological parameters as predicted by the model under consideration. The constraints are obtained by performing a statistical analysis based on the Markov Chain Monte Carlo (MCMC) method (Section \ref{cosmo-const}), and considering three sets of observational data: the Hubble parameter ($H(z)$), the Type Ia supernova sample (Pantheon+SH0ES), and the redshift distortion sample ($f\sigma_8$). We find that, although posing constraints on the model parameters presents some difficulties when individual datasets are considered, the joint statistical analyses allow one to set strong constraints on the parameters such that the model fits the data accurately, within the observational uncertainties. In addition, our model predictions for the considered cosmological parameters are found to be consistent with those reported by Planck or DESI (within the $\sim 1\sigma$ C.L.). Remarkably, we obtain a present value of the Hubble parameter, $H_0$, laying between the values reported by Planck \cite{Planck:2018vyg} and SH0ES \cite{Riess:2021jrx,Riess:2022mme}, alleviating the tension between these observations.

Our results also indicate that the value of the deviation parameter $b$ that best fits the data (Section \ref{cosmo-const}) is larger than expected, considering the perturbative approach implemented in \cite{Oliveros:2023ouq} to find the approximate solution. The impact of such a large value reflects on the redshift evolution of the cosmological parameters $w_{\rm{eff}}$, $q$, $w_{\rm{DE}}$ and $\Omega_{\rm{DE}}$ presented in Section \ref{sec_cosmo-dynamics}, with particular impact on $w_{\rm{DE}}$, from which an oscillatory evolution at late times is obtained. 

We also present results from statistical (Section \ref{sec_IC}) and dynamics (\ref{sec_cosmo-dynamics}) tests to verify the validity of the model under consideration, and address the conclusions in Section \ref{conclude}.
\section{$f(R)$ Gravity: Preliminaries}\label{model} 
\noindent In general, the gravitational action of $f(R)$ gravity in the presence  of matter components is given by
\begin{equation}\label{eq1}
S=\int{d^4x\sqrt{-g}\left(\frac{f(R)}{2\kappa^2}+\mathcal{L}_{\rm{M}}\right)},
\end{equation}
where $g$ denotes the determinant of the metric tensor $g^{\mu\nu}$, $\kappa^2=8\pi G=1/M_{\rm{p}}^2$, with $G$ being the Newton's constant and $M_{\rm{p}}$ the reduced Planck mass. $\mathcal{L}_{\rm{M}}$ represents the Lagrangian density for the matter components (relativistic and non-relativistic perfect matter fluids). The term $f(R)$ is for now an arbitrary function of the Ricci scalar $R$. Variation with respect to the metric gives the equation of motion
\begin{equation}\label{eq2}
f_R(R)R_{\mu\nu}-\frac{1}{2}g_{\mu\nu}f(R)+(g_{\mu\nu}\square-\nabla_\mu\nabla_\nu)f_R(R)=\kappa^2T_{\mu\nu}^{(\rm{M)}},
\end{equation}
where $f_R\equiv \frac{df}{dR}$, $\nabla_{\mu}$ is the covariant derivative associated with the Levi-Civita connection of the metric, and $\square\equiv \nabla^\mu\nabla_\mu$. In addition, $T_{\mu\nu}^{(\rm{M})}$ is the matter energy--momentum tensor which is assumed to be a perfect fluid. Considering the flat Friedman-Robertson-Walker (FRW) metric,
\begin{equation}\label{eq3}
ds^2=-dt^2+a(t)^2\delta_{ij}dx^idx^j,
\end{equation}
with $a(t)$ representing the scale factor, the time and spatial components of Eq.~(\ref{eq2}) are given, respectively, by
\begin{equation}\label{eq4}
3H^2f_R=\kappa^2(\rho_{\rm{m}}+\rho_{\rm{r}})+\frac{1}{2}(Rf_R-f)-3H\dot{f}_R,
\end{equation}
and
\begin{equation}\label{eq5}
-2\dot{H}f_R=\kappa^2\left(\rho_{\rm{m}}+\frac{4}{3}\rho_{\rm{r}}\right)+\ddot{f}_R-H\dot{f}_R,
\end{equation}
where $\rho_{\rm{m}}$ is the matter density and $\rho_{\rm{r}}$ denotes the density of radiation. The over-dot denotes a derivative with respect to the cosmic time $t$ and $H\equiv \dot{a}/a$ is the Hubble parameter. Note that in the spatially flat FLRW Universe, the Ricci scalar $R$ takes the form
\begin{equation}\label{eq5'}
R=6(2H^2+\dot{H}).
\end{equation}
If there is no interaction between non-relativistic matter and radiation, then these components obey separately the conservation laws:
\begin{equation}\label{eq6}
\dot{\rho}_{\rm{m}}+3H\rho_{\rm{m}}=0,\quad \dot{\rho}_{\rm{r}}+4H\rho_{\rm{r}}=0.
\end{equation}
As usual in the literature, it is possible to rewrite the field equations (\ref{eq4}) and (\ref{eq5}) in the Einstein-Hilbert form:
\begin{equation}\label{eq7}
3H^2=\kappa^2\rho,
\end{equation}
\begin{equation}\label{eq8}
-2\dot{H}^2=\kappa^2(\rho+p),
\end{equation}
where $\rho=\rho_{\rm{m}}+\rho_{\rm{r}}+\rho_{\rm{DE}}$ and $p=p_{\rm{m}}+p_{\rm{r}}+p_{\rm{DE}}$ correspond to the total effective energy density and total effective pressure density of the cosmological fluid. In this case, the dark energy component has a geometric origin, and after some manipulation in Eqs.~(\ref{eq4}) and (\ref{eq5}), we obtain the effective dark energy and pressure corresponding to $f(R)$-theory given by
\begin{equation}\label{eq9}
\rho_{\rm{DE}}=\frac{1}{\kappa^2}\left[\frac{Rf_R-f}{2}+3H^2(1-f_R)-3H\dot{f}_R\right],
\end{equation}
and
\begin{equation}\label{eq10}
p_{\rm{DE}}=\frac{1}{\kappa^2}[\ddot{f}_R-H\dot{f}_R+2\dot{H}(f_R-1)-\kappa^2\rho_{\rm{DE}}].
\end{equation}
It is easy to show that $\rho_{\rm{DE}}$ and $p_{\rm{DE}}$ defined in this way satisfy the usual energy conservation equation
\begin{equation}\label{eq11}
\dot{\rho}_{\rm{DE}}+3H(\rho_{\rm{DE}}+p_{\rm{DE}})=0;
\end{equation}
in this case we assume that the equation of state (EoS) parameter for this effective dark energy satisfies the relation
$w_{\rm{DE}}=p_{\rm{DE}}/\rho_{\rm{DE}}$, and in explicit form it is given by
\begin{equation}\label{eq12}
w_{\rm{DE}}=-1-\frac{H\dot{f}_R+2\dot{H}(1-f_R)-\ddot{f}_R}{\frac{1}{2}(f_RR-f)-3H\dot{f}_R+3(1-f_R)H^2}.
\end{equation}
In the following sections, our analysis will be focused on the $f(R)$ gravity model, defined by
\begin{equation}\label{eq13}
f(R)=R-2\,\Lambda\,e^{-(b\Lambda/R)^{n}},
\end{equation}
where $b$ and $n$ are positive real dimensionless parameters, and $\Lambda$ is the cosmological constant. This model was introduced in Ref. \citep{Oliveros:2023ouq}, and is a reparameterization of a specific viable $f(R)$ gravity model studied in \citep{Oliveros:2023ewl,Granda:2020afq}. Furthermore, it is shown that the HS model is a limiting case of this model. In the literature, other authors have studied some $f(R)$ gravity models with exponential functions of the scalar curvature (see, for instance, Refs.~\citep{Linder:2009jz, Odintsov:2007zu, Odintsov:2017qif, Odintsov:2018qug, Odintsov:2023nqe}). As will become clear in the next sections, the two free parameters ($n$ and $b$) can be used to better fit the observational data and reduce the current observed tensions for some cosmological parameters. Additionally, this model could be extended by adding a term proportional to $R^2$ in Eq.~(\ref{eq13}), allowing the inflationary period to be considered, thereby strengthening the constraints on the model. 

From the specific form of this model, and as has been demonstrated in Ref.~\citep{Basilakos:2013nfa}, it is possible to derive an analytic approximation for the expansion rate $H(z)$. This approximate analytical expression was found by one of the authors in Ref. \citep{Oliveros:2023ouq}, and it is given by
\begin{equation}\label{eq_Hz_fR}
\begin{aligned}
E^2(z) &\equiv \frac{H^2(z)}{H_0^2}=1-\Omega_{m0}+(1+z)^3\Omega_{m0}\\
&+\frac{6b(\Omega_{m0}-1)^2\left(-4+\Omega_{m0}(9-3\Omega_{m0}+z(3+z(z+3))
(1+(3+2z(3+z(z+3)))\Omega_{m0}))\right)}{(4+(-3+z(3+z(z+3)))\Omega_{m0})^3}\\
&+\frac{b^2(\Omega_{m0}-1)^3}{(1+z)^{24}\left(\frac{4(1-\Omega_{m0})}{(1+z)^3}+\Omega_{m0}\right)^8}
\bigg[5120(\Omega_{m0}-1)^6+9216(1+z)^3(\Omega_{m0}-1)^5\Omega_{m0} \\
&- 30144(1+z)^6(\Omega_{m0}-1)^4\Omega_{m0}^2 + 31424(1+z)^9(\Omega_{m0}-1)^3\Omega_{m0}^3-9468(1+z)^{12} (\Omega_{m0}-1)^2\Omega_{m0}^4  \\
&- 4344(1+z)^{15}(\Omega_{m0}-1)\Omega_{m0}^5+\frac{37}{2}(1+z)^{18}\Omega_{m0}^6\bigg],
\end{aligned}
\end{equation}
where for simplicity, it has been assumed that $\Omega_{r0} = 0$, $n = 1$ and made the substitution $N=-\ln{(1+z)}$.

\section{Cosmological constraints}\label{cosmo-const}
This section is devoted to the description of the statistical analysis performed and the considered observational data, to obtain constraints on the free parameter of the model, $b$, as well as on some of the relevant cosmological parameters, as predicted by the $f(R)$ model. We also present a comparison with the predictions of the $\Lambda$CDM model when the same statistical analysis and datasets are considered.

The statistical analysis used here is based on the well known Markov Chain Monte Carlo (MCMC) method implemented with the \texttt{emcee} package \citep{Foreman-Mackey:2012any} to find the parameters that maximize a user-defined likelihood function
\begin{equation}\label{eq_Likelihood}
    \mathcal{L}(D|z;\bm{\theta}) = -\ln p(D|z;\bm{\theta}) = -\frac12 \chi^2(z;\bm{\theta}), 
\end{equation}
where $D$ refers to the analyzed dataset(s), $\bm{\theta}$ is the vector of free the parameters to fit (the actual elements of this vector depend on the dataset under consideration, as explained in the following sections), and $z$ is the independent variable which, for our case corresponds to the redshift. 

For each dataset considered in this paper (the Hubble parameter, the Type Ia supernova --Pantheon+-- and the redshift space distortion (RSD)), a suitable $\chi^2$ function is defined, considering the particular number of data and the observed uncertainties. In addition, combinations of the different datasets are also considered, looking for strengthen the constraints on the relevant parameters, in which case the corresponding $\chi^2$ function would be the sum of the individual functions for each dataset, i.e.,
\begin{equation}
    \chi_{\rm{tot}}^2(z;\bm{\theta}) = \sum_i\chi_i^2(z;\bm{\theta}), \quad i = (H(z), {\rm Pantheon}, f\sigma_8).
\end{equation}

Using the MCMC method benefits by including previously known information about the parameters. This is done by adding a set of suitable \emph{priors} that makes the \texttt{emcee} package to explore the parameters within a defined range, with a specified probability distribution. In order to avoid any possible bias on the analysis, flat priors are enforced for all parameters, with the corresponding ranges shown in table \ref{tab_priors}.
\begin{table}[width=0.75\linewidth,cols=6,pos=h]
\caption{Defined ranges for the parameters to fit, included as flat priors in the MCMC analysis.}
    \label{tab_priors}%
\begin{tabular*}{\tblwidth}{@{} LCCCCC@{} }
\toprule
    Parameter & $H_0$ $^a$ & $\Omega_{m0}$ & $M$        & $\sigma_8$ & $b$ \\ 
    \midrule
    Range     & (60, 80)              & (0.1, 0.4)    & (-22, -16) & (0.6, 1.0) & (-2, 2) \\
    \bottomrule
\multicolumn{6}{l}{\footnotesize{$^a$ $H_0$ is measured in km/s/Mpc.}}
\end{tabular*}
\end{table}

As can be seen in Table \ref{tab_priors}, we allow $H_0$ to float in an interval that includes the two values that are currently under discussion given the independent measurements by Planck, $H_0^{\rm{Planck}} = [67.36\pm0.54]$ km s$^{-1}$ Mpc$^{-1}$ \cite{Planck:2018vyg} and SH0ES, $H_0^{\rm{SH0ES}} = [73.30\pm1.04]$ km s$^{-1}$ Mpc$^{-1}$ \citep{Riess:2021jrx} (see also Ref.~\citep{Riess:2022mme} for a reported value with reduced uncertainty). In doing so, we are able to test whether our model shows indications of alleviating the tension between the two observations. Moreover, it is noteworthy that we have considered a range that includes negative values for $b$, even though the imposed conditions of the current model, i.e., $f_{R}>0$ and $f_{RR}>0$ for $R>R_0$ $(>0)$ (where $R_0$ is the Ricci scalar at the present time), and also $0<\frac{Rf_{RR}}{f_{R}}(r)<1$ at the de Sitter point, $r=-\frac{R f_R}{f}=-2$, require that $b > 0$; allowing the fitter to explore beyond this condition guarantees that the possibility that $b=0$ is correctly considered.

\subsection{The Hubble parameter data}\label{sec_Hz}
For the observational Hubble parameter, we consider the data reported in \citep{Cao:2021uda} based on cosmic chronometers (CC) and radial Baryon Acoustic Oscillations (BAO) methods (see Ref.~\citep{Cao:2021uda} for additional details). In this case, the $\chi^2$ used for the likelihood maximization is defined as
\begin{equation}\label{eq_chi2_Hz}
    \chi^2_{Hz}(z;\bm{\theta}) = \sum_{k=1}^{N_d}
    \frac{\left(
    H_{{\rm fR},k}(z;\bm{\theta}) - H_{{\rm obs},k}(z)
    \right)^2}{\sigma_k^2},
\end{equation}
with $N_d = 40$, and $\bm{\theta} = (H_0, \Omega_{m0}, b)$. Here, $H_{{\rm fR},k}$ and $H_{{\rm obs},k}$ are the $f(R)$-model prediction (Eq.~(\ref{eq_Hz_fR})), and the observed values of the expansion rate, respectively, and $\sigma_k$ is the corresponding observational error. An analogous $\chi^2$ function is used to test the prediction of the $\Lambda$CDM model for $H(z)$, for which the parameter vector reduces to $\bm{\theta}_{\Lambda\rm{CDM}} = (H_0, \Omega_{m0})$.

\begin{figure}
\centering
    \includegraphics[width=0.75\columnwidth]{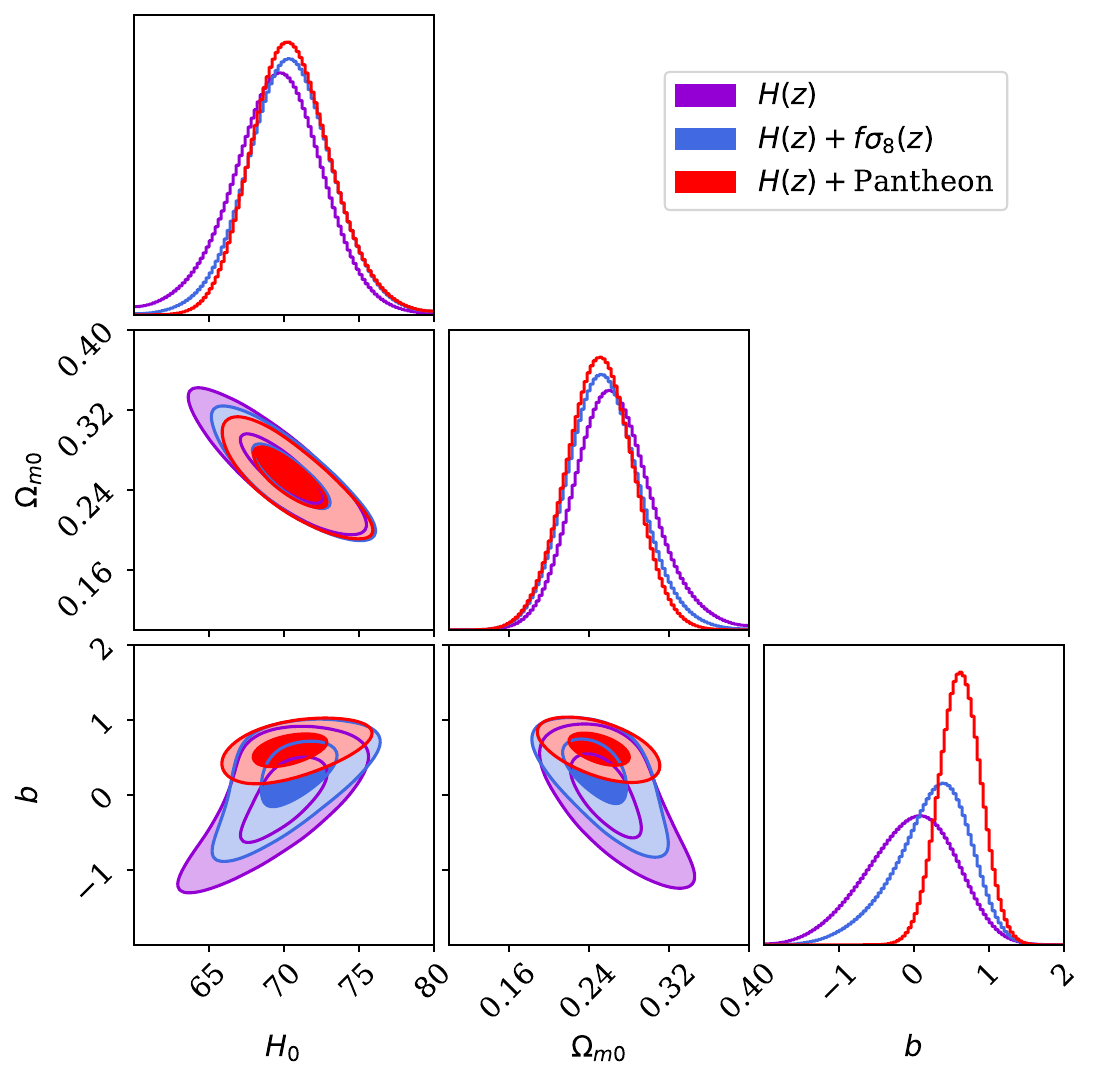}
    \caption{Contour plots and 1-D posterior probabilities obtained from the MCMC analysis of the $H(z)$ (purple) and its combination with $f\sigma_8(z)$ (blue) and Pantheon (red, for which the column corresponding to $M$ was cut -see next section-) observational data, for the parameters $\left(H_0,\Omega_{m0},b\right)$.
    }
    \label{fig_cornerCombo_HzmSN}
\end{figure}

The results of this fit are shown in Fig.~\ref{fig_cornerCombo_HzmSN}, where the $1\sigma$-- and $2\sigma$--confidence contour plots and posterior probabilities for the considered parameters are exhibited (purple contours and lines). The best-fit (BF) values of the parameters are also presented in Table \ref{tab_MCMCintervals}, where we also write our results obtained from fitting the $\Lambda$CDM model to this dataset, and the values reported by the Planck \citep{Planck:2018vyg} and DESI \citep{DESI:2024mwx} Collaborations.

Regarding the present value of the Hubble parameter, $H_0$, within the $1\sigma$ interval, the prediction of our model lies well between the Planck and SH0ES observations, closer to the former. Our model prediction for $\Omega_{m0}$ agrees with the Planck value within a ${\sim}1.6\sigma$ C.L. Note that, although the parameter $b$ is not strongly constrained by this dataset and the best fit is negative, the prediction is consistent with $b=0$, and the allowed interval expands up to $b\lesssim 0.8$ at $\,\sim2\sigma$ C.L.

\subsection{The standardized distance modulus - Type Ia Supernova Sample}\label{sec_SN}
For the Ia supernova distance modulus we consider the Pantheon + SH0ES (referred to as Pantheon here) database described in Refs.~\citep{Riess:2021jrx,Brout:2022vxf}, comprising 1701 data points in a range of $0.001 \leq z \leq 2.3$. The analysis was performed with a suited $\chi^2$ function, considering both statistical and systematic uncertainties through a covariance matrix, $\bm{C}_{\rm{cov}}$:
\begin{equation}\label{eq_chi2_muSN}
    \chi^2_{\rm{Pantheon}} = \left[\mu_{\rm{fR}}(z;\bm{\theta}) - \mu_{\rm{obs}}(z)\right]^{T} \bm{C}_{\rm{cov}}^{-1} \, \left[\mu_{\rm{fR}}(z;\bm{\theta}) - \mu_{\rm{obs}}(z)\right].
\end{equation}
Both, the covariance matrix and the observed distance modulus $\mu_{\rm{obs}}$, were obtained from the Pantheon + SH0ES data release \citep{PantheonPlusSH0ES_2022}. For the model prediction, we have
\begin{equation}\label{eq_muSN}
    \mu_{\rm{fR}}(z;M,\bm{\tilde{\theta}}) = M + 25 + 5 \log_{10}D_L(z;\bm{\tilde{\theta}}),
\end{equation}
where $M$ is the absolute magnitude, the parameters vector is $\bm{\tilde{\theta}} = (H_0, \Omega_{m0}, b)$, and $D_L(z;\bm{\theta})$ is the luminosity distance given by
\begin{equation}
    D_L(z;\bm{\tilde{\theta}}) = \frac{c\,(1+z)}{H_0} \int_0^z \frac{dz'}{E(z';\bm{\tilde{\theta}})},
\end{equation}
where $c$ is the speed of light and, as before, $E(z';\bm{\tilde{\theta}})$ is given by Eq.~(\ref{eq_Hz_fR}).

Notice that, in Eq.~(\ref{eq_muSN}), the distance modulus is written as a function of the absolute magnitude, $M$, in order to make it clear that, for our analyses where Pantheon data are included, we also include $M$ as a parameter to fit, i.e., $\bm{\theta} = (M, H_0, \Omega_{m0}, b)$ in the corresponding $\chi^2$ function, like Eq.~(\ref{eq_chi2_muSN}). As for the analysis of the expansion rate in the previous section, here we also carry out the fit of the prediction of the $\Lambda$CDM model for the distance modulus with the same $\chi^2$ function, Eq.~(\ref{eq_chi2_muSN}), reducing the vector of parameters to $\bm{\theta} = (M, H_0, \Omega_{m0})$.

The contour plots, together with the posterior probabilities for the fitted parameters for the $f(R)$ model, are shown in Fig.~\ref{fig_cornerCombo_HzmSN2}. As expected, the absolute magnitude $M$ is well constrained by this dataset and the prediction of the $f(R)$ model (see Table \ref{tab_MCMCintervals}) is consistent with the value reported in Ref.~\citep{Riess:2021jrx}. A strong correlation between some of the parameters is evident in the contours obtained, in particular for ($M, H_0$) and ($\Omega_{m0}, b$), exhibiting long allowed areas, covering most of the range of values studied.

It is interesting to note that the Pantheon data alone are not enough to constrain the present value of the Hubble expansion rate $H_0$, an aspect also observed, for instance, by the authors of Ref.~\citep{Brout:2022vxf}. Furthermore, although the best fit for $b$ is rather large, the data analysis is statistically consistent with $b = 0$ (see Table \ref{tab_MCMCintervals}). In addition, the observed correlation between $\Omega_{\rm{m}0}$ and $b$, makes $\Omega_{\rm{m}0}$ agree with the Planck and DESI observations for $b\sim 0$, as expected.

\subsection{Combining the Hubble expansion rate and the Pantheon+SH0ES datasets}\label{sec_Hz_n_SN}
\begin{figure}
\centering
    \includegraphics[width=0.75\columnwidth]{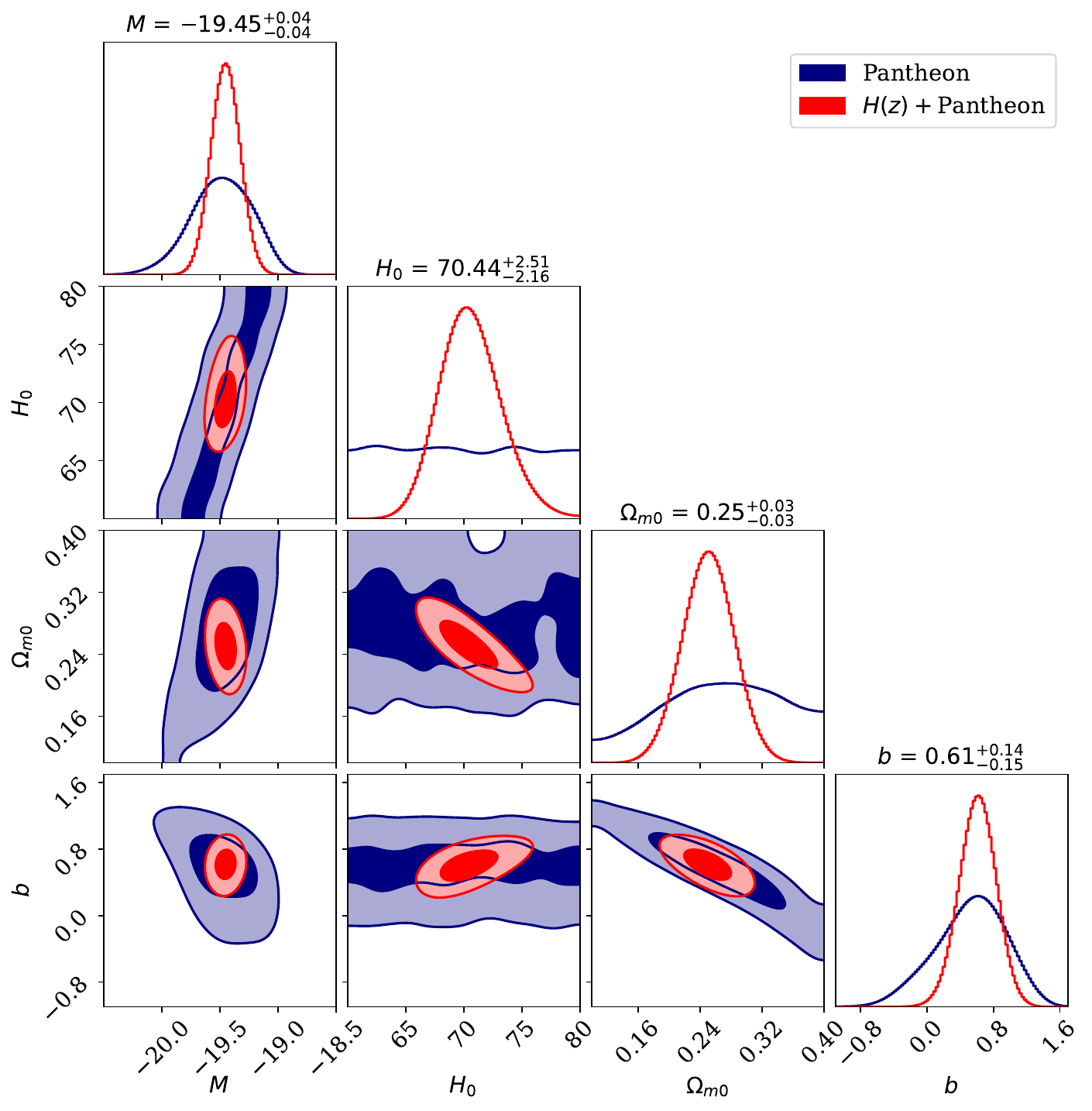}
    \caption{Contour plots and 1-D posterior probabilities obtained from the MCMC analysis of the Pantheon (dark blue) observational data, as well as its combination with $H(z)$ datasets (red), for the parameters $\left(M, H_0,\Omega_{m0},b\right)$. The numbers over the 1-D posteriors correspond to the joint analysis.}
    \label{fig_cornerCombo_HzmSN2}
\end{figure}

As anticipated at the beginning of this section, a joint analysis was also implemented considering the two previously described datasets, adding the corresponding $\chi^2$-functions, Eqs.~(\ref{eq_chi2_Hz}) and (\ref{eq_chi2_muSN}). The combined fit produces the expected results, shown as (red) contours and 1-D posterior probabilities in Figs.~\ref{fig_cornerCombo_HzmSN} and \ref{fig_cornerCombo_HzmSN2} (where the column for $M$ is included), and in Table \ref{tab_MCMCintervals} (see the row for $H(z) + $SN). Despite the larger number of data available from the Pantheon sample, it is the combination with the CC data that breaks the degeneracy between the parameters, resulting in stringent constraints; this is particularly apparent from the 1-D histograms for $H_0$ and $\Omega_{\rm{m}0}$. For these quantities, the joint fit keeps the model prediction close to that obtained from the $H(z)$ dataset alone (see also Fig.~\ref{fig_cornerCombo_HzmSN}), but enhancing the corresponding limits (i.e.~reducing the allowed regions).

This last feature is exceptionally noticeable for the parameter $b$, for which not only we obtain a rather large best-fit value ($b_{\rm{BF}} = 0.614$), but also $b=0$ is predicted to be excluded at more than $3\sigma$ C.L. Although one would expect the deviation parameter $b$ to be close to zero, bringing our model close to $\Lambda$CDM, as will become clear later (Sec.~\ref{sec_comparison}), the proposed $f(R)$-model with a large deviation parameter successfully fit the observational data considered. In addition, these results are consistent with earlier studies \citep{Basilakos:2013nfa}, where values of $b$ of the order $\mathcal{O}(1)$ are also obtained and are compatible with the conclusions proposed by the authors of Ref.~\citep{Odintsov:2024woi}.

\subsection{The redshift space distortion, $f\sigma_8$ - The growth Sample}\label{sec_fs8}
The last dataset considered here is the value of the growth rate $f(z)$ multiplied by the amplitude of the matter power spectrum on the scale of $8h^{-1}\,\,\rm{Mpc}$, $\sigma_8(z)$, usually written as $f\sigma_8(z)$. This quantity is considered the best observable to distinguish between modified gravity theories (such as $f(R)$ gravity models) and $\Lambda$CDM, since many $f(R)$ gravity models are virtually indistinguishable from the $\Lambda$CDM model at the background level \citep{Cardone:2012xv}. We consider a total of $N_d = 26$ data points for different redshifts, $0.013 \leq z \leq 1.944$ \citep{Nesseris:2017njc,Saridakis:2021xqy,Alestas:2022gcg}, with the $\chi^2$ function defined as
\begin{equation}\label{eq_chi2_fs8}
    \chi^2_{f\sigma_8} = \sum_{k = 1}^{N_d} 
    \frac{\left[(f\sigma_8)_{\rm{fR},k}(z;\bm{\theta}) - (f\sigma_8)_{\rm{obs},k}(z)\right]^2}{\sigma_k^2}.
\end{equation}

As in the previous cases, $\sigma_k$ is the corresponding error for each observational value $(f\sigma_8)_{\rm{obs},k}(z)$, which is compared against the model prediction, $(f\sigma_8)_{\rm{fR},k}(z;\bm{\theta})$, with $\bm{\theta} = (\sigma_8, H_0, \Omega_{m0}, b)$. The predicted growth rate is computed though the following relation \citep{Nesseris:2017njc,Mhamdi:2024kgu}:
\begin{equation}\label{eq_fs8_def}
    (f\sigma_8)_{\rm{fR}}(z;\bm{\theta}) = \sigma_8 \frac{\delta_{\rm{m}}'(z;\bm{\theta})}{\delta_{\rm{m}}(z=0)},
\end{equation}
where $\sigma_8=\sigma_8(z=0)$, $\delta_{\rm{m}}\equiv \delta\rho_{\rm{m}}/\rho_{\rm{m}}$ is the gauge-invariant matter density perturbation (the density contrast), and the prime stands for the derivative with respect to the redshift, $z$. Clearly, to obtain the theoretical prediction from Eq.~(\ref{eq_fs8_def}), it is necessary to calculate $\delta_{\rm{m}}$. The equation governing the evolution of this quantity for the $f(R)$ gravity has been derived previously in the literature, considering the subhorizon approximation ($k^2/a^2\gg H^2$) \citep{Tsujikawa:2007gd, Tsujikawa:2009ku}, and it is written as
\begin{equation}\label{eq_delta_m_t}
\ddot{\delta}_{\rm{m}}+2H\dot{\delta}_{\rm{m}}-4\pi G_{\rm{eff}}(a,k)\rho_{\rm{m}}\delta_{\rm{m}}=0;
\end{equation}
here the dot denotes the differentiation with respect to the cosmic time, $G_{\rm{eff}}(a,k)$ is the effective gravitational ``constant'', $k$ is the comoving wave number, $a$ is the scale factor normalized to unity at present epoch, and $\rho_{\rm{m}}$ is the background matter density. In order to facilitate our calculations, we rewrite Eq.~(\ref{eq_delta_m_t}) in terms of $z$, as follows:
\begin{equation}\label{eq_delta_m}
 \delta_{\rm{m}}''(z) + \left(\frac{E^{2\,\prime}(z)}{2E^{2}(z)}-\frac{1}{1+z}\right)\delta_{\rm{m}}'(z)-\frac{3\Omega_{\rm{m0}}}{2E^{2}(z)}(1+z)\frac{G_{\rm{eff}}(z,k)}{G_{\rm{N}}}\delta_{\rm{m}}(z)=0;
\end{equation}
in this case, the explicit form for $G_{\rm{eff}}(z,k)$ is
\begin{equation}\label{eq_G_eff}
 G_{\rm{eff}}(z,k)=\frac{G_{\rm{N}}}{f_R}\left[1+\frac{k^2(1+z)^2(f_{RR}/f_R)}{1+3k^2(1+z)^2(f_{RR}/f_R)}\right],
\end{equation}
where $G_{\rm{N}}$ is the Newton constant. Equation (\ref{eq_delta_m}) has been expressed in terms of $E^2(z)$, since this function is known in an explicit form in our case. To solve (\ref{eq_delta_m}) numerically, we adopt initial conditions for the density contrast, and its first derivative that are consistent with those observed at very high redshifts (matter era), matching that of the $\Lambda$CDM model.

Statistical analysis allows us to set constraints on the parameters $\bm{\theta} = (\sigma_8, H_0, \Omega_{m0}, b)$. However, as observed in Table \ref{tab_MCMCintervals} (see the row for $f(R) - f\sigma_8$), the allowed interval obtained for $H_0$ is considerably large; in fact, the constraints on this parameter are rather weaker than for the other cases, indicating that these datasets alone are not enough to provide a robust fit. 

To overcome this situation, we performed a series of joint fits that combine the growth sample with 
\begin{itemize}
    \item the Hubble expansion rate sample, $\bm{\theta} = (\sigma_8, H_0, \Omega_{m0}, b)$,
    \item the Pantheon sample, $\bm{\theta} = (M, \sigma_8, H_0, \Omega_{m0}, b)$,
    \item both Hubble expansion rate and Pantheon samples at the same time, $\bm{\theta} = (M, \sigma_8, H_0, \Omega_{m0}, b)$.
\end{itemize}
\begin{figure}
\centering
    \includegraphics[width=0.75\columnwidth]{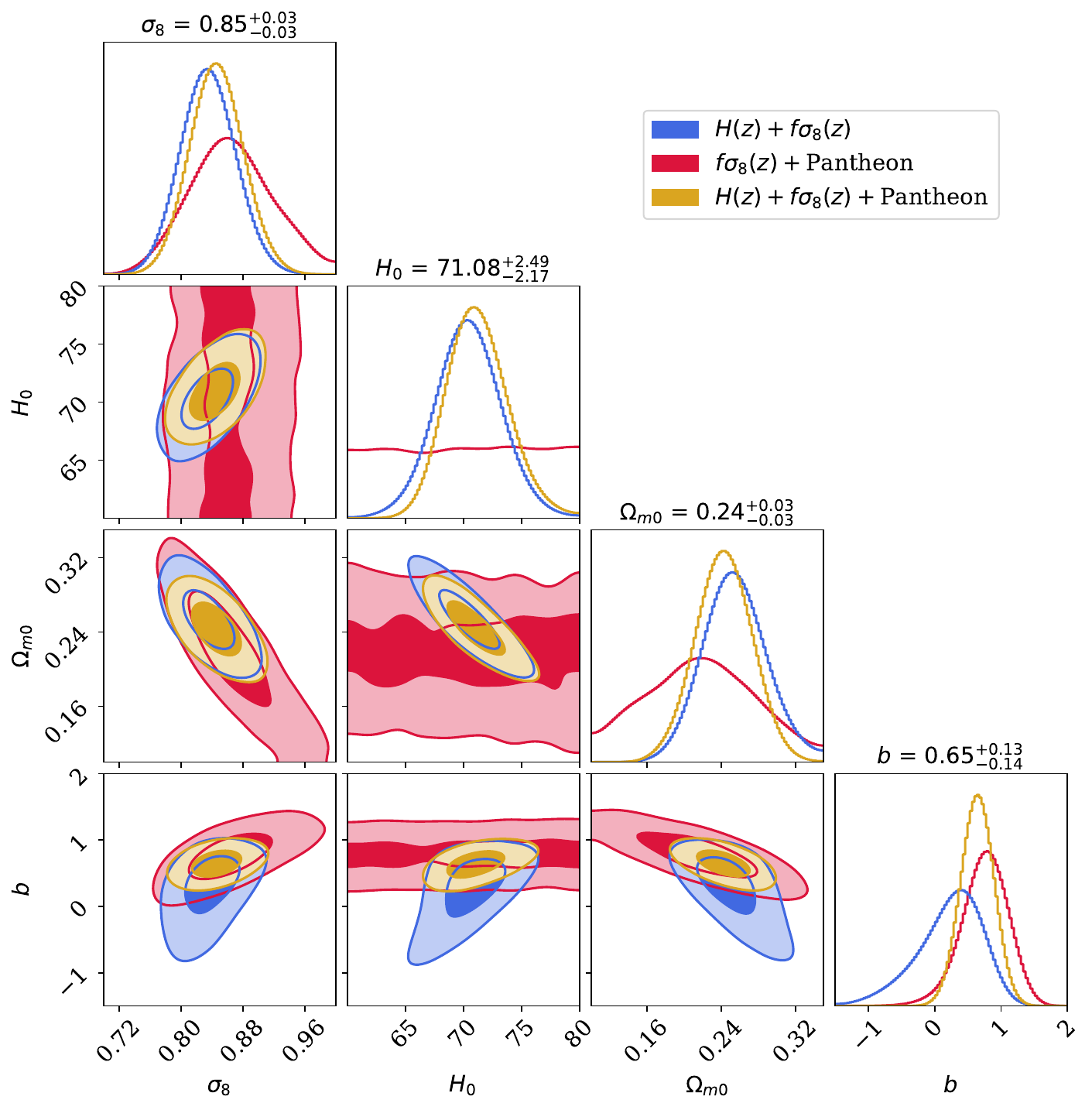}
    \caption{Contour plots and 1-D posterior probabilities obtained from the MCMC analysis, for the combination of $H(z)$, $f\sigma_8(z)$, and Pantheon observational data for the parameters $\left(\sigma_8,H_0,\Omega_{m0},b\right)$. For a better comparison, here the column corresponding to $M$ has been removed from the analyses including the Pantheon sample.}
    \label{fig_cornerCombo_Hzfs8mSN}
\end{figure}

The corresponding posterior distributions for the considered parameters resulting from these three different analyses are shown in Fig.~\ref{fig_cornerCombo_Hzfs8mSN}, where the numbers on top of each column correspond to the inferred values from the combination of the three datasets (gold 2D contours and histograms), and the allowed regions for $M$ are not displayed to facilitate comparison. Table \ref{tab_MCMCintervals} also shows the resulting intervals at a 68\% C.L. for all the fits.

Looking at Fig.~\ref{fig_cornerCombo_Hzfs8mSN} one can notice the apparent effect of the different datasets combination. As neither Pantheon nor the growth rate sample independently is sufficient to set constraints on $H_0$, the corresponding joint analysis does not perform better and a correlation between $\sigma_8$ and $\Omega_{m0}$ persists (also observed when the growth rate sample is analyzed alone). It is the inclusion of the CC dataset that breaks all degeneracies, so that all the parameters are better bounded, with the most evident impact observed for $H_0$, for which the combined data give a well-constrained allowed region around $H_0 = 71.1$ km/s/Mpc. Interestingly, the bounds of $\sigma_8$ are also improved with the combination, showing that, although this parameter is undoubtedly not limited by the $H(z)$ data, the joint fit with Pantheon and the growth rate samples makes the model predict a smaller allowed region for $\sigma_8$, rejecting values outside $\sim(0.81, 0.87)$ (see bottom left panel of Fig.~\ref{fig_cornerCombo_Hzfs8mSN}).

We stress here that, for the case of the $(\sigma_8,\Omega_{\rm{m}0})$ space, a similar result was obtained by the authors of Ref.~\citep{Nesseris:2017njc}, in a context where variants of the Hu-Sawicki model were studied. Comparing this 2D-parameter space with a more recent study \citep{Hou:2024blc}, where constraints from the redshift-space galaxy skew spectra are set for some cosmological parameters (although not in the context of $f(R)$ gravity models), we see that we obtain compatible results both, at the 2D-contour level, and the allowed region for each parameter. This provides an interesting insight about the possibility of strengthening our constraints even further by the inclusion of non-Gaussian information of the cosmic large-scale structure, a task which might be considered in a future work.

\begin{figure}
\centering
    \includegraphics[width=0.75\columnwidth]{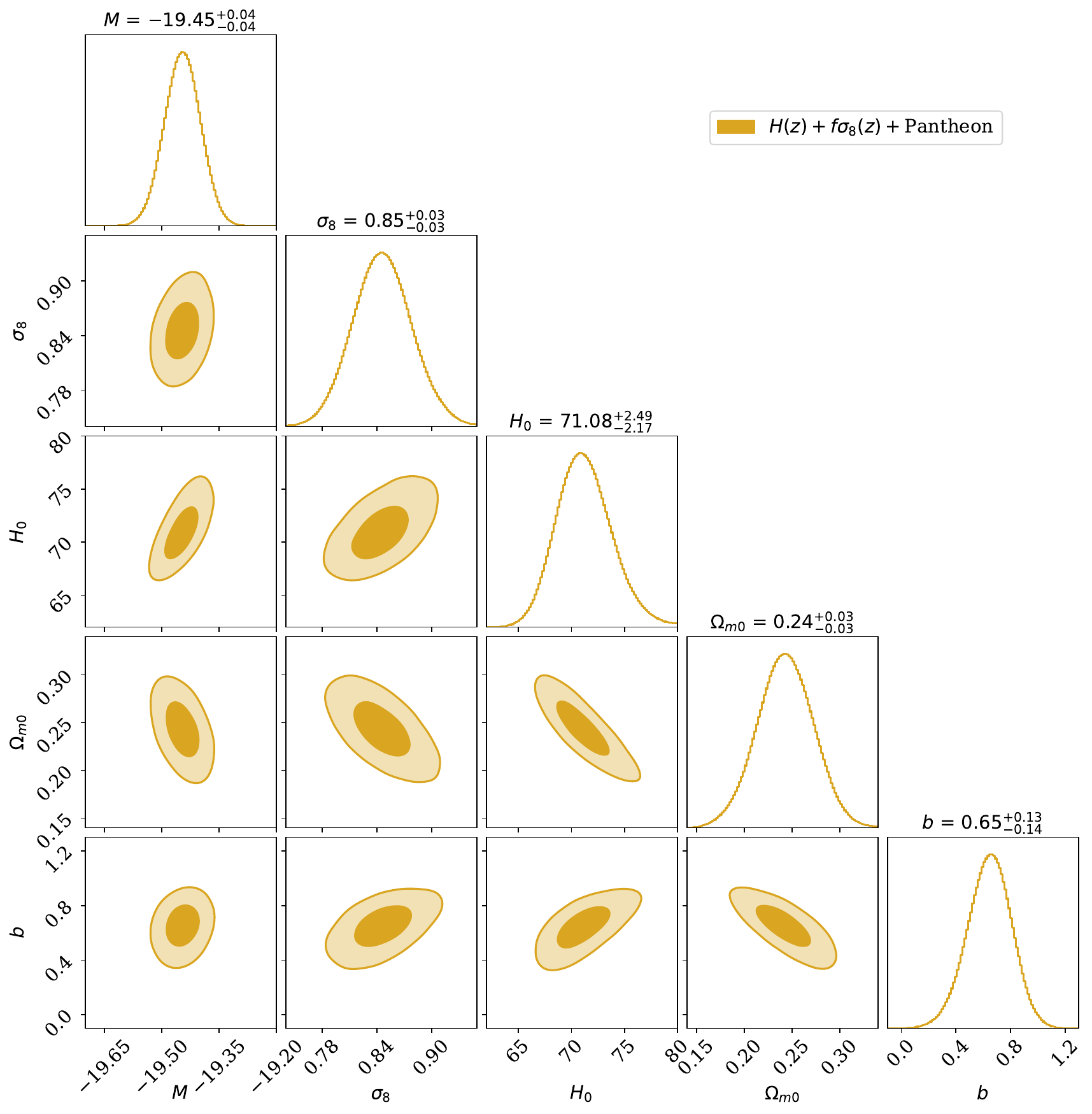}
    \caption{Contour plots and 1-D posterior probabilities obtained from the MCMC analysis, for the combination of the three datasets considered in this work, for the parameters $\left(M, \sigma_8,H_0,\Omega_{m0},b\right)$. The results for $M$ are included for completness.}
    \label{fig_cornerAll}
\end{figure}
\begin{figure}
\centering
    \includegraphics[width=0.70\columnwidth]{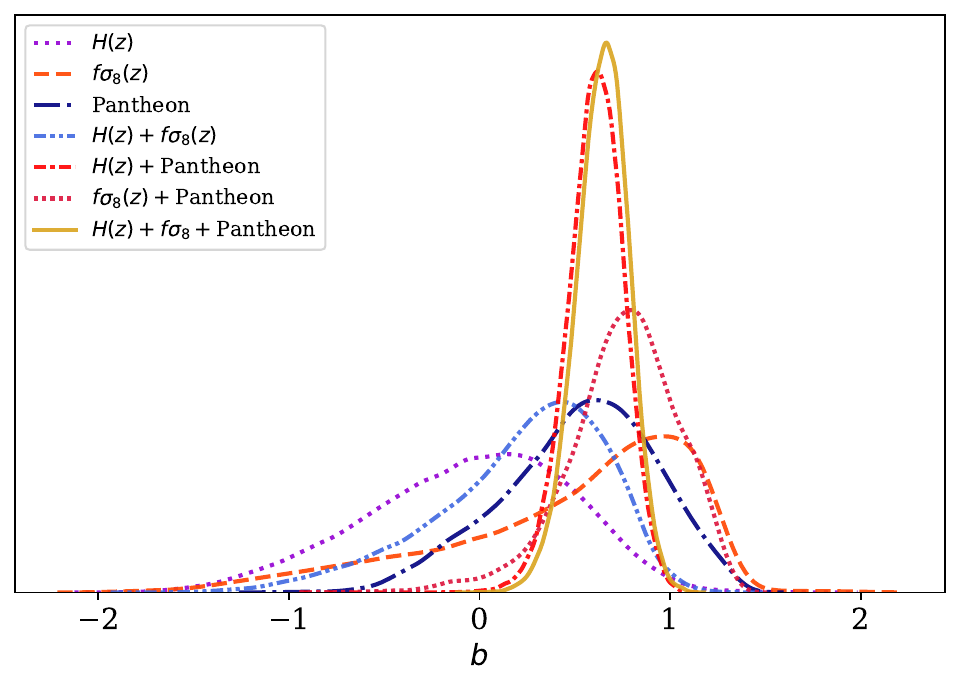}
    \caption{The $b$ 1-D posterior probabilities obtained from the MCMC analyses of the $H(z)$, $f\sigma_8(z)$, and Pantheon observational data.}
    \label{fig_posteriors2}
\end{figure}

Another relevant aspect to stress here are the results for the perturbative parameter, $b$. As noted in the previous section (see Fig.~\ref{fig_cornerCombo_HzmSN}, and Tab.~\ref{tab_MCMCintervals}), when only the $H(z)$ sample is considered, the allowed region for $b$ goes to lower values, including $b=0$; however, when the other two samples are considered in the analysis (see also Fig.~\ref{fig_cornerAll}), not only the constraints are strengthened, but also $b$ moves up to rather large values, now excluding $b = 0$ at $\gtrsim 3.5\sigma$. These large values for $b$ might not be expected, considering its perturbative nature, but are not totally unanticipated, since similar results have been observed before \citep{Basilakos:2013nfa,Sultana:2022qzn} (see also \citep{Nesseris:2017njc}, where a particular degenerate hypergeometric model was considered, obtaining a best fit of $b \approx 6$), and, as mentioned before, is consistent with the conclusions of recent work by Odintsov and collaborators \citep{Odintsov:2024woi}.

\begin{table}[width=1.0\linewidth,cols=7,pos=h]
\caption{Resulting $1\sigma$ allowed intervals for the fitted parameters from the MCMC analysis. The values in the first (second) row correspond to those for the $\Lambda$CDM model from Planck 2018 \citep{Planck:2018vyg} (DESI 2024 \citep{DESI:2024mwx}); the following rows are the result of fitting the $\Lambda$CDM model predictions to the corresponding dataset, while the last part of the table shows the corresponding predictions from our $f(R)$ model. Note that, here, $\rm{SN}$ refers to the Pantheon dataset, and by ``\emph{All}'' we mean the combination of the three datasets.}
    \label{tab_MCMCintervals}%
\begin{tabular*}{\tblwidth}{@{} LLCCCCC@{} }
\toprule
    \multicolumn{2}{c}{Model} & $H_0$ $^a$ & $\Omega_{m0}$ & $\sigma_{8}$ & $M$ & $b$ \\ \hline
    $\Lambda$CDM & Planck \citep{Planck:2018vyg} & $67.4\pm0.5$ & $0.315\pm0.007$ & $0.811\pm0.006$ & -- & -- \\
    \midrule
    $\Lambda$CDM & DESI \citep{DESI:2024mwx} & $68.52\pm0.62$ $^b$ & $0.295\pm0.015$ $^c$ & $0.8135\pm0.0053$ $^d$ & $-19.253\pm0.027$ & -- \\
    \midrule
    \multirow{7}{2em}{$\Lambda$CDM} & $H(z)$ & $73.1_{-2.8}^{+2.7}$ & $0.237_{-0.025}^{+0.029}$ & -- & -- & -- \\ 
    & $\rm{SN}$ & $70.0 \pm 6.8$ & $0.361 \pm 0.018$ & -- & $-19.34_{-0.22}^{+0.20}$ & -- \\ 
    & $H(z)+\rm{SN}$ & $68.6 \pm 1.4$ & $0.338 \pm 0.017$ & -- & $-19.51\pm0.04$ & -- \\
    & $f\sigma_8$ & -- & $0.294_{-0.043}^{+0.046}$ & $0.806_{-0.031}^{+0.035}$ & -- & -- \\ 
    & $H(z)+f\sigma_8$ & $71.6 \pm 2.4$ & $0.253_{-0.023}^{+0.026}$ & $0.832 \pm 0.028$ & -- & -- \\
    & ${\rm SN}+f\sigma_8$ & $70.0 \pm 6.8$ & $0.353 \pm 0.017$ & $0.777 \pm 0.022$ & $-19.34_{-0.22}^{+0.20}$ & -- \\
    & \emph{All} & $65.2_{-1.3}^{+1.4}$ & $0.333 \pm 0.016$ & $0.786\pm 0.022$ & $-19.50\pm0.04$ & -- \\ 
    \midrule
    \multirow{7}{2em}{$f(R)$} & $H(z)$ & $69.6_{-2.8}^{+2.6}$ & $0.263_{-0.031}^{+0.036}$ & -- & -- & $-0.013_{-0.596}^{+0.495}$ \\ 
    & {\rm SN} & $69.8_{-6.8}^{+6.9}$ & $0.269_{-0.084}^{+0.079}$ & -- & $-19.48_{-0.23}^{+0.24}$ & $0.560_{-0.448}^{+0.359}$ \\ 
    & $H(z)+{\rm SN}$ & $70.4_{-2.2}^{+2.5}$ & $0.251\pm+0.028$ & -- & $-19.45\pm0.04$ & $0.614_{-0.149}^{+0.138}$ \\
    & $f\sigma_8$ & $70.0_{-6.7}^{+6.8}$ & $0.242_{-0.082}^{+0.083}$ & $0.846_{-0.054}^{+0.071}$ & -- & $0.644_{-0.883}^{+0.432}$ \\ 
    & $H(z)+f\sigma_8$ & $70.4\pm2.5$ & $0.254_{-0.029}^{+0.032}$ & $0.834_{-0.030}^{+0.032}$ & -- & $0.297_{-0.528}^{+0.354}$ \\
    & ${\rm SN} +f\sigma_8$ & $70.2_{-6.9}^{+6.7}$ & $0.219_{-0.061}^{+0.062}$ & $0.863_{-0.049}^{+0.056}$ & $-19.55_{-0.23}^{+0.24}$ & $0.772_{-0.283}^{+0.263}$ \\
    & \emph{All} & $71.1_{-2.2}^{+2.5}$ & $0.243_{-0.026}^{+0.027}$ & $0.845 _{-0.029}^{+0.030}$ & $-19.45\pm0.04$ & $0.649_{-0.142}^{+0.128}$ \\ 
\bottomrule
    \multicolumn{7}{l}{\footnotesize{$^a$}$H_0$ is measured in km/s/Mpc.}\\
    \multicolumn{7}{l}{\footnotesize{$^b$}DESI BAO.}\\
    \multicolumn{7}{l}{\footnotesize{$^c$}DESI BAO + CMB.}\\
    \multicolumn{7}{l}{\footnotesize{$^d$}DESI BAO + Planck[plik] + CMB lensing.}\\
\end{tabular*}
\end{table}

Let us conclude this section by pointing out that although it would be natural to expect $b$ to be close to zero, our statistical analyses indicate that this is not quite the case for the considered model and data samples; this is nicely evidenced in Fig.~\ref{fig_posteriors2}, where the posterior probabilities for the parameter $b$ resulting from the different analyses are shown. However, one must notice that the large value obtained for the perturbative parameter is statistically strong and the fit to the data is compensated (see Sec.~\ref{sec_comparison}) by the other relevant cosmological parameters considered in the analysis, $(M, H_0, \Omega_{\rm{m}0},\sigma_8)$,  which are different from those reported by Planck and DESI (see Table \ref{tab_MCMCintervals}), within the $\Lambda$CDM model. Indeed, when we perform the same statistical analysis implemented for our $f(R)$ model to the $\Lambda$CDM predictions, the results appear to be more compatible between these two models (see the $\Lambda$CDM and $f(R)$ sets of rows in Table \ref{tab_MCMCintervals}), with a remarkable distinction coming from the fact that our proposed $f(R)$ model (i.e., with $b \neq 0$) predicted $H_0$ value lies in between the reported measurements by Planck and SH0ES, albeit closer to the later than to the former.

\subsection{Model predictions vs. observational data}\label{sec_comparison}
As an important evaluation of the results presented in the previous sections, the obtained values for the restricted parameters ($H_0, \Omega_{\rm{m}0}, M, \sigma_8, b$) are used to draw the evolution of the Hubble expansion rate, the distance modulus Ia-SN, and the space distortion, in terms of the redshift, $z$, as predicted by our $f(R)$ model. 

\begin{figure}
\centering
    \includegraphics[width=0.75\columnwidth]{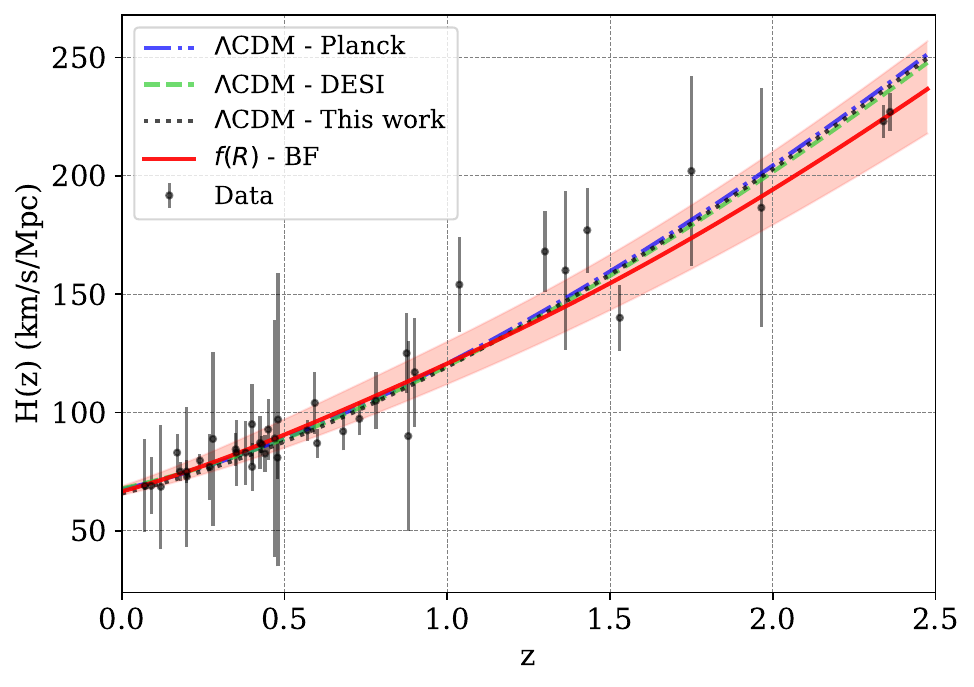}
    \caption{Evolution of the Hubble parameter with the redshift, $z$, as predicted for the $f(R)$ model presented in this work (dot-dashed red), compared against observational data (black dots with the vertical lines indicating the uncertainty). The prediction by the $\Lambda$CDM model (full blue for Planck and dashed black for DESI) is also shown for comparison.}
    \label{fig_hz_dataCompare}
\end{figure}

\begin{figure}
\centering
    \includegraphics[width=0.75\columnwidth]{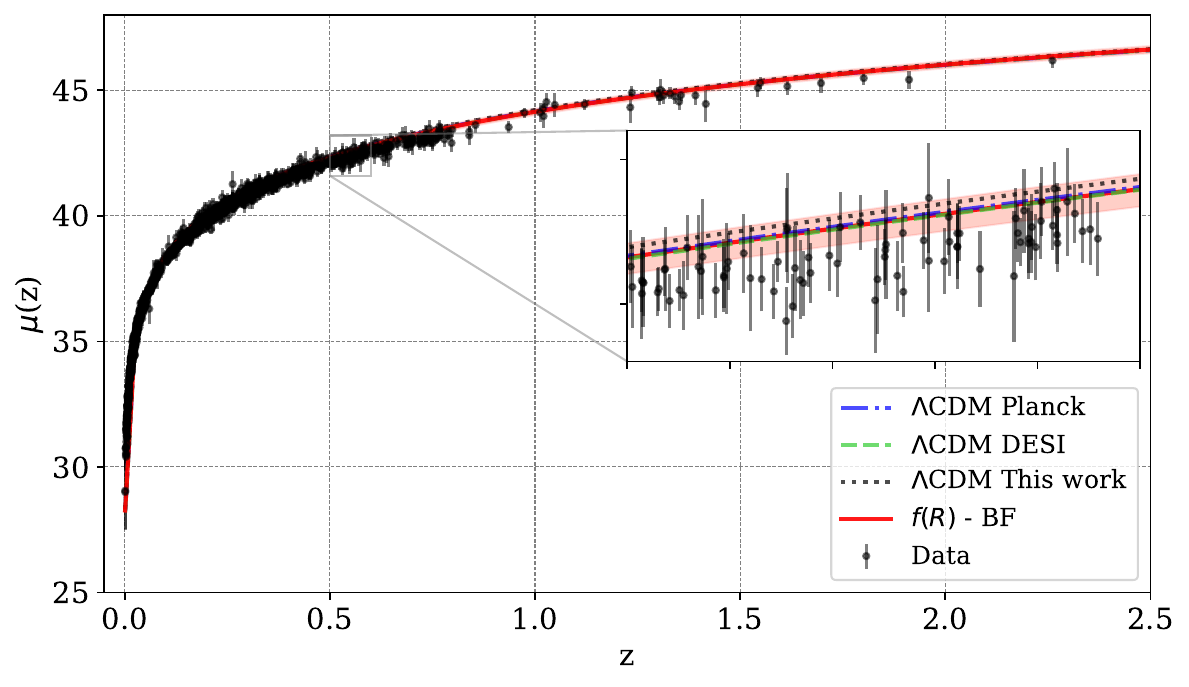}
    \caption{As in Fig.~\ref{fig_hz_dataCompare}, but for the distance modulus as a function of the redshift.}
    \label{fig_mbz_dataCompare}
\end{figure}

\begin{figure}
\centering
    \includegraphics[width=0.75\columnwidth]{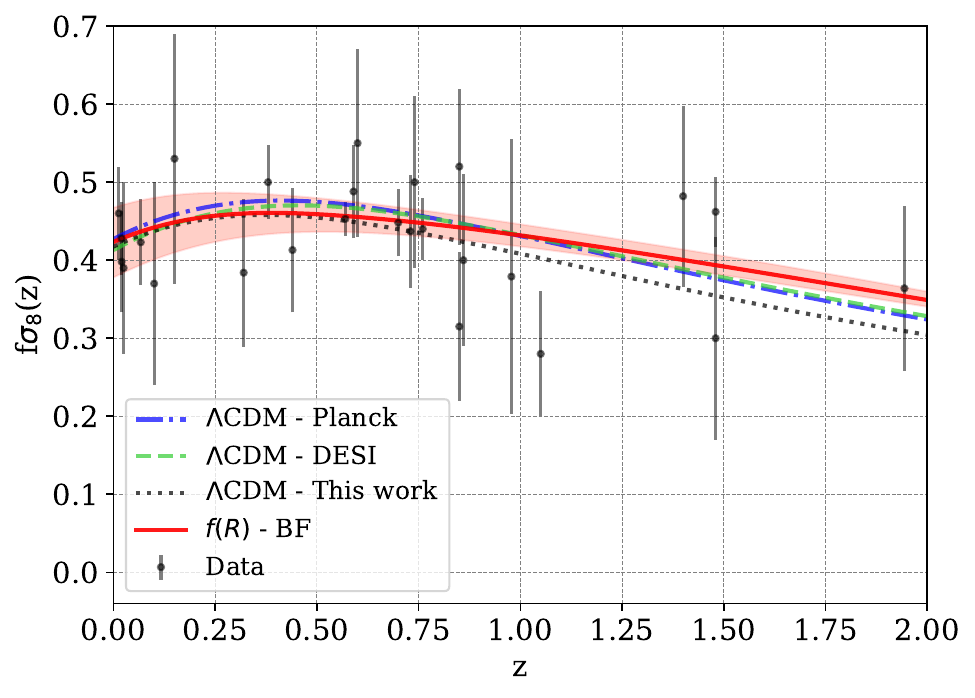}
    \caption{As in Fig.~\ref{fig_hz_dataCompare}, but for the growth rate as a function of the redshift.}
    \label{fig_fs8z_dataCompare}
\end{figure}

This is shown in Figs.~\ref{fig_hz_dataCompare}-\ref{fig_fs8z_dataCompare}, where the $f(R)$ predictions (red dashed-dotted line) are compared with each published data sample (black dots with the vertical line indicating the corresponding data uncertainty), as well as with the predictions from the $\Lambda$CDM model, considering both Planck \citep{Planck:2018vyg} (blue line) and DESI \citep{Riess:2021jrx} (green dashed line) reported observations.

For a complete and more consistent comparison, in Figs.~\ref{fig_hz_dataCompare}-\ref{fig_fs8z_dataCompare} we also included our own fit of the predictions of the $\Lambda$CDM model to the considered data samples (black dotted line). Although not easily visible in Fig.~\ref{fig_mbz_dataCompare}, the three figures also exhibit a light red band obtained by allowing the parameters to move up to the $1\sigma$ allowed limits. It is evident that our model very well reproduces the observations and that, despite the large value of the perturbative parameter $b$, the proposed model does not deviate considerably from $\Lambda$CDM.

\begin{table}[width=.9\linewidth,cols=7,pos=h]
\caption{Results of the information criteria analyses comparing the $\Lambda$CDM model with the $f(R)$ model proposed in this work, with the different sets of considered data samples.}\label{tab_InfoCrit}
\begin{tabular*}{\tblwidth}{@{} LLCCCCC@{} }
\toprule
        Dataset & Model & $\chi^2_{\rm{min}}$ & AIC & $|\Delta$AIC$|$ & BIC & $|\Delta$BIC$|$ \\ 
        \midrule
        \multirow{2}{4em}{$H(z)$} &
        $\Lambda$CDM & 18.70 & 25.36 & 0    & 29.76 & 0    \\
         & $f(R)$    & 18.87 & 28.02 & 2.64 & 33.63 & 3.87 \\ 
        \midrule
        \multirow{2}{5em}{Pantheon} &
        $\Lambda$CDM & 1752.51 & 1758.52 & 0     & 1774.83 & 0     \\
         & $f(R)$    & 1839.81 & 1847.83 & 89.31 & 1869.56 & 94.74 \\ 
        \midrule
        \multirow{2}{8em}{$H(z)$ + Pantheon} &
        $\Lambda$CDM & 1781.93 & 1787.94 & 8.73 & 1804.32 & 3.28 \\
         & $f(R)$    & 1771.20 & 1779.22 & 0    & 1801.04 & 0    \\ 
        \midrule
        \multirow{2}{4em}{$f\sigma_8$} &
        $\Lambda$CDM & 14.92 & 19.44 & 0    & 21.44 & 0    \\
         & $f(R)$    & 13.58 & 20.67 & 1.23 & 23.36 & 1.92 \\ 
        \midrule
        \multirow{2}{6em}{$H(z) + f\sigma_8$} &
        $\Lambda$CDM & 34.85 & 41.23 & 0      & 47.42 & 0    \\
         & $f(R)$    & 33.51 & 42.17 & 0.93   & 50.27 & 2.85 \\ 
        \midrule
        \multirow{2}{8em}{$f\sigma_8$ + Pantheon} &
        $\Lambda$CDM & 1769.20 & 1777.23 & 0     & 1799.02 & 0     \\
         & $f(R)$    & 1828.89 & 1838.93 & 61.70 & 1866.16 & 67.14 \\ 
        \midrule
        \multirow{2}{8em}{\emph{All}} &
        $\Lambda$CDM & 1797.69 & 1805.71 & 10.90 & 1827.59 & 5.43 \\
         & $f(R)$    & 1784.78 & 1794.81 & 0     & 1822.16 & 0    \\
    \bottomrule
    \end{tabular*}
\end{table}

\subsection{Information Criteria}\label{sec_IC}
In this section we implement a different evaluation of the fits described in the previous sections, using two standard information criteria (IC): the Akaike Information Criterion (AIC) and the Bayesian Information Criterion (BIC). This procedure provides a way to compare a set of model with their predictions given by datasets (see Ref.~\citep{Mandal:2023cag} and references therein for a complete description, and Refs.~\citep{Sultana:2022qzn,Nesseris:2017vor}, where this analysis is also implemented). This analysis is useful to compare models with different number of parameters and the number of data points for the different data samples considered.

Specifically, the AIC estimator is given by \citep{Mandal:2023cag}
\begin{equation}\label{eq_AIC}
    \rm{AIC} = -2\ln(\mathcal{L}_{\rm{max}}) + 2k + \frac{2k(k+1)}{N_{\rm{tot}} - k - 1},
\end{equation}
while the BIC evidence estimator is computed through
\begin{equation}\label{eq_BIC}
    \rm{BIC} = -2\ln(\mathcal{L}_{\rm{max}}) + k\log(N_{\rm{tot}}),
\end{equation}
where $k$ is the number of free parameters in the proposed model, $\mathcal{L}_{\rm{max}}$ is the maximum likelihood value of the dataset(s) considered for analysis, and $N_{\rm{tot}}$ is the number of data points. Then, to compare the models, we compute the relative differences between the IC,
\begin{equation}\label{eq_deltaIC}
    \Delta IC_{\rm{model}} = IC_{\rm{model}} - IC_{\rm{min}},
\end{equation}
where $IC_{\rm{min}}$ is the minimum value of IC of the set of competing models \citep{Mandal:2023cag}. According to the authors of Ref.~\citep{Mandal:2023cag}, a value $\Delta IC \leq 2$ indicates the statistical compatibility of the compared models; obtaining $2 < \Delta IC < 6$ points to a moderate tension between the models, and $\Delta IC \geq 10$ hints towards a strong tension. In general, the larger $\Delta IC_{\rm{model}}$, the stronger the evidence against the model compared to the model with $IC_{\rm{min}}$.

The results of the IC analysis are presented in Table \ref{tab_InfoCrit}. Although we only use two models ($\Lambda$CDM vs.~$f(R)$), the comparison is made from the results of the statistical analyses of the different datasets (separately and jointly), as described above. For each case, in Table \ref{tab_InfoCrit} we report the values of $\chi^2_{\rm{min}}$, AIC (Eq.~(\ref{eq_AIC})), BIC (Eq.~(\ref{eq_BIC})), and $|\Delta IC|$ (computed for both criteria using Eq.~(\ref{eq_deltaIC})).

If only $\chi^2_{\rm{min}}$ is considered, we see that compared to $\Lambda$CDM, the proposed $f(R)$ model provides a better fit to the RSD sample and to the combined analyses (except when the RSD sample is combined with the Pantheon dataset). Then, by looking at the AIC and BIC values, for which the number of parameters is considered, the situation is more convoluted, as the IC is lower for $\Lambda$CDM in some cases and larger in others. In particular, note that the largest differences ($|\Delta IC|$) are obtained when the Pantheon sample is used in the statistical analysis (either alone or in combination with $H(z)$ or $f\sigma_8$). Although the combination of $H(z)$ and Pantheon datasets points to a mild preference for our proposed model over $\Lambda$CDM, Pantheon alone and Pantheon$+ f\sigma_8$ strongly prefer $\Lambda$CDM.

The results of the other analyses do not provide a compelling indication in favor of any of the models, but point to the compatibility between them and to the fact that both models are equally likely to reproduce the data (though it is interesting to point that the combination of the three datasets gives $|\Delta$AIC$|>10$ in favor of the $f(R)$ model). Also, one has to consider the fact that the proposed $f(R)$ model originates as a perturbation of $\Lambda$CDM, so the results are not astonishing \footnote{Let us point out that, as marked in \citep{Nesseris:2017njc} and detailed in \citep{Nesseris:2012cq}, this kind of analysis should not be taken as a final word when comparing different models, but as a complementary tool.}, and additional tests might be performed.

\section{Cosmological dynamics in late-time}\label{sec_cosmo-dynamics} 
Finally, setting the parameters of the model to the BF values obtained from the joint fit (Table \ref{tab_MCMCintervals}, last row), we can take a look at the cosmological dynamics at late time as described by the $f(R)$ model studied here. 

\subsection{Om($z$) Diagnostic}\label{sec_OmDiagnostic}
An interesting tool to study the dynamics of a particular model is the Om diagnostic proposed in Ref.~\citep{Sahni:2006pa}, which relies on the Hubble parameter, $H(z)$. With this diagnostic, it is also possible to analyze the differences between the proposed model and $\Lambda$CDM. The diagnostic is performed by computing
\begin{equation}\label{eq_OmD}
    \rm{Om}(z) = \frac{E^2(z) - 1}{(1+z)^3 -1},
\end{equation}
where $E^2(z) = H(z)/H_0$. Looking at the evolution of $\rm{Om}(z)$, one can obtain information about the nature of DE as predicted by the considered model: if the model predicts a quintessence behavior, $\rm{Om}(z)$ would exhibit a negative slope (decreasing evolution); if, instead, the prediction favors a phantom DE, $\rm{Om}(z)$ increases with $z$, showing a positive slope; finally, $\rm{Om}(z)$ remains constant, corresponds to a cosmological constant DE, \emph{i.e.}, the standard $\Lambda$CDM model.

For the $f(R)$ model studied in this work, we can calculate $\rm{Om}(z)$ using the analytical solution, Eq.~(\ref{eq_Hz_fR}), considering the BF values of the relevant parameters $(\Omega_{\rm{m}0},b)$, obtained from the joint statistical analysis of the $H(z)$+Pantheon+$f\sigma_8$ data (last row of Table \ref{tab_MCMCintervals}). 
\begin{figure}
\centering
    \includegraphics[width=0.75\columnwidth]{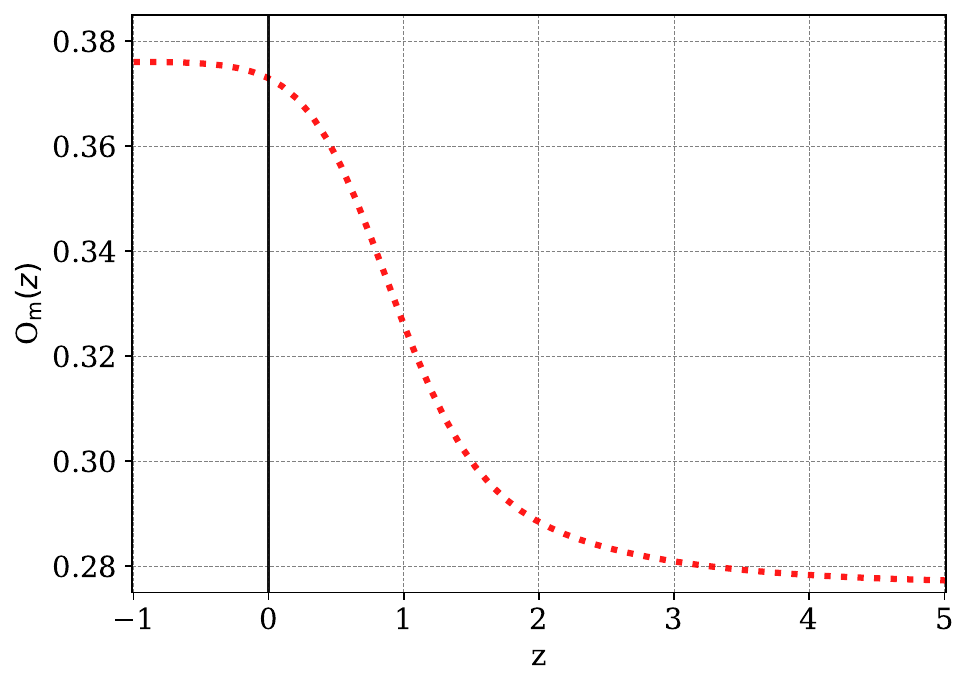}
    \caption{Evolution of Om terms of the redshift, as predicted by the $f(R)$ model proposed in this work, considering the constraints from $H(z)$, Pantheon and $f\sigma_8$ samples.}
    \label{fig_fR_Om}
\end{figure}

The resulting evolution of $\rm{Om}(z)$ is shown in Fig.~\ref{fig_fR_Om}. Notice how, for $z < 0$ and $z \gtrsim 3.5$, $\rm{Om}(z)$ presents a negligible variation (zero slope), indicating that the effective DE would behave like a cosmological constant. For the region in between, and for $z=0$, in particular, $\rm{Om}(z)$ decreases (negative slope), implying that the effective DE of our model displays a quintessence-like behavior, which is consistent with the evolution of the DE EoS, $w_{\rm{DE}}$ at most of the corresponding $z$ interval (see left bottom panel of Fig.~\ref{fig_fR_cosmoParams}). 

\subsection{Cosmological parameters}\label{sec_CosmoDynamics}
\begin{figure}
\centering
    \includegraphics[width=1.0\textwidth]{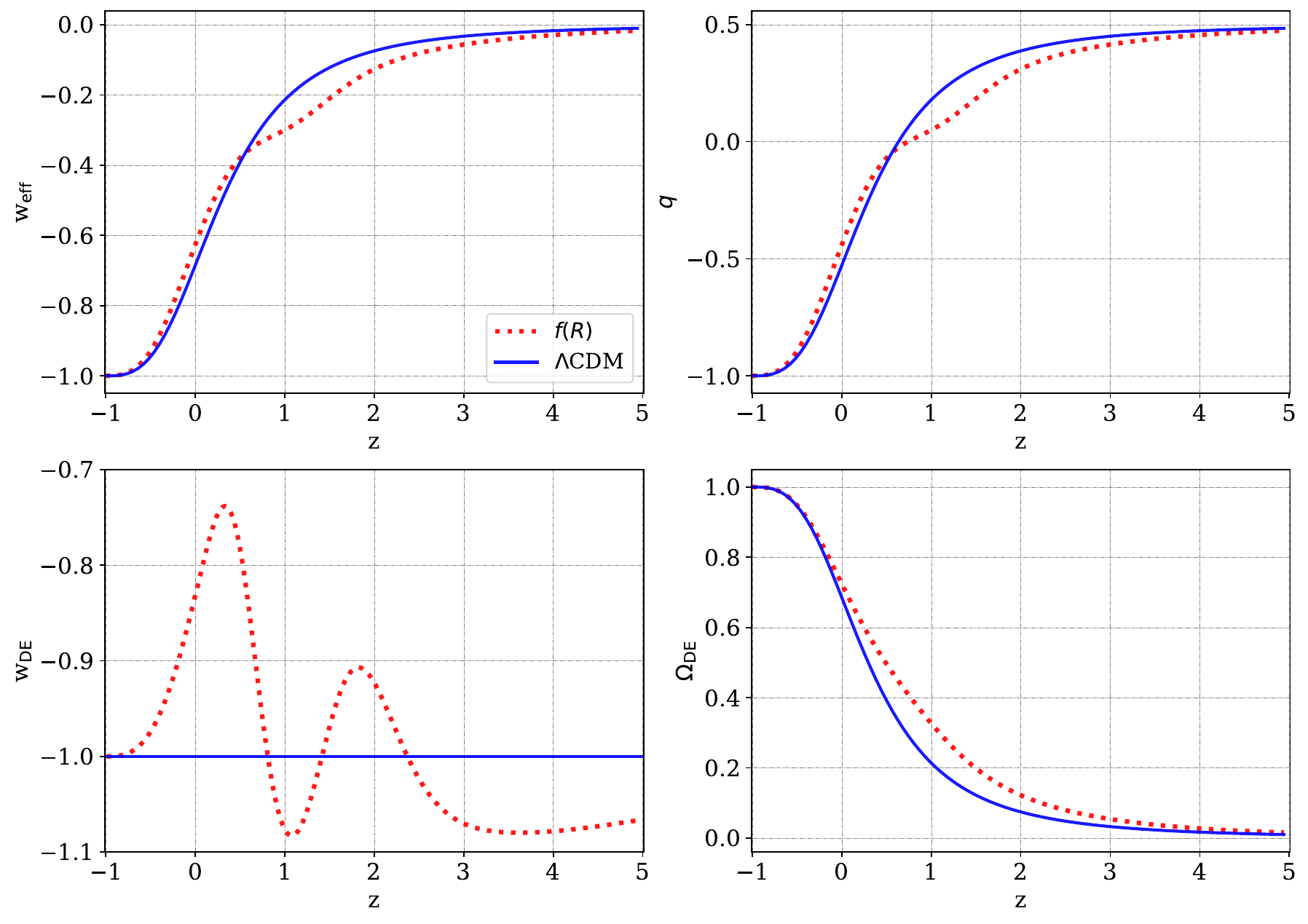}
    \caption{Comparison of the evolution of some cosmological parameters in terms of the redshift, as predicted by the $\Lambda$CDM (blue-full line) model and the $f(R)$ (red-dotted line) model proposed in this work.}
    \label{fig_fR_cosmoParams}
\end{figure}
We now consider interesting cosmological parameters as given by the proposed $f(R)$ model, which provide insights on the predictions and evolution of the model, as well as a suitable way to compare with $\Lambda$CDM. In particular, here we examine the effective EoS, 
\begin{equation}\label{eq_w_eff}
w_{\rm{eff}}=-1+\frac{1}{3}(1+z)\frac{(E^2(z))'}{E^2(z)},
\end{equation}
the deceleration parameter, 
\begin{equation}\label{eq_q}
q=-1+\frac{1}{2}(1+z)\frac{(E^2(z))'}{E^2(z)},
\end{equation}
the DE EoS, 
\begin{equation}\label{eq_w_DE}
w_{\rm{DE}}=-1+\frac{1}{3}(1+z)\frac{(\rho_{\rm{DE}}(z))'}{\rho_{\rm{DE}}(z)},
\end{equation}
and the DE density, 
\begin{equation}\label{eq_Omega_DE}
\Omega_{\rm{DE}}=1-\frac{\Omega_{\rm{m}0}(1+z)^3}{E^2(z)};
\end{equation}
in all the above expressions, the prime indicates derivative with respect to $z$.

The evolution of these quantities predicted by the $f(R)$ model is shown in Fig.~\ref{fig_fR_cosmoParams} (red dotted lines) in terms of the redshift, $z$, where we also include the prediction of $\Lambda$CDM (blue lines), for comparison. 

Despite the fact that the statistical analysis showed a preference for $b \sim \mathcal{O}(10^{-1})$, the cosmological evolution of $w_{\rm{eff}}$, $q$ and $\Omega_{\rm{DE}}$ predicted by the $f(R)$ model closely resembles the prediction of $\Lambda$CDM. The largest deviation appears in the range $0.5 \lesssim z \lesssim 3$, most certainly due to the fact that the approximated solution implemented in this analysis considers a perturbative expansion up to a second order; if additional terms (proportional to $b^n$, $n>2$) had been considered, the $f(R)$ model would have resulted to be much closer to $\Lambda$CDM, and the red dotted lines in Fig.~\ref{fig_fR_cosmoParams} would be almost indistinguishable from the blue ones. This is expected since it has already been shown in Refs.~\citep{Granda:2020afq, Oliveros:2023ewl} that using the exact (numerical) solution for the Hubble expansion rate, the $f(R)$ model is essentially indistinguishable from $\Lambda$CDM at the background level.

As observed in the bottom left panel of Fig.~\ref{fig_fR_cosmoParams}, $w_{\rm{DE}}$ shows a considerable deviation from $w_{\rm{DE}} = -1$ along the depicted range, especially for $z \lesssim 4$, where oscillatory behavior is observed. This discrepancy (although with a lower amplitude) was already anticipated in Ref.~\citep{Oliveros:2023ouq} for a smaller value of the deviation parameter $b$; it is also apparent that $w_{\rm{DE}} \to -1$ at the early stages of the Universe ($z \gtrsim 4$), as also observed in \citep{Oliveros:2023ouq}. This is another indication of the effect of using the perturbative expansion up to the second order. In fact, it is reasonable to think that, as a consequence of $b \sim \mathcal{O}(10^{-1})$ as obtained from the analysis of the observational data, additional terms in the expansion of Eq.~(\ref{eq_Hz_fR}), proportional to $b^3$ and larger powers, could contribute substantially to the solution, likely mitigating the oscillations of $w_{\rm{DE}}$. It is also important to note that this oscillatory evolution has already been observed by other authors, for instance, in the context of modified gravity models \citep{Granda:2020ikv,Granda:2022ttn}, or considering dynamical dark energy models \citep{Zhao:2017cud,Escamilla:2024fzq}.

\section{Conclusions}\label{conclude} 
In this paper, we have performed a statistical analysis of a known viable $f(R)$ gravity model that includes an exponential function of the scalar curvature, Eq.~(\ref{eq13}), with a specific parameter $b$ governing its deviation from $\Lambda$CDM. In this context, we implemented the approximate analytical solution for the expansion rate, $E^2(z)$, shown in Eq.~(\ref{eq_Hz_fR}), from which some observational quantities can be computed, allowing to investigate the impact of truncating the perturbative expansion with respect to $b$.

In addition to $H(z)$, we considered the distance modulus, $\mu(z)$, and the growth rate multiplied by the amplitude of the matter power spectrum at $8h^{-1}\rm{\,\,Mpc}$, $f\sigma_8(z)$. Hence, for the statistical analysis we used observational data from cosmic chronometers and radial Baryon Acoustic Oscillations methods (Section \ref{sec_Hz}), the SN Ia Pantheon+SH0ES sample (Section \ref{sec_SN}), and the growth sample (Section \ref{sec_fs8}). We individually analyzed these data samples to set constraints on the model parameters $\left(M,\sigma_8,H_0,\Omega_{m0},b\right)$ and found that both the SN Ia and the growth sample individually do not provide reasonable bounds on $H_0$, and that, from the three separate analyses, the value of the deviation parameter $b$ that best fits each data set is $\sim\mathcal{O}(10^{-1})$, and that $b = 0$ is not statistically excluded (see the first, second and fourth rows of Table \ref{tab_MCMCintervals}).

Strengthened constraints on the parameters were obtained by performing joint analyses. By only combining $H(z)$ and the Pantheon samples, the bounds on $\left(H_0,\Omega_{m0},b\right)$ are considerably improved (Fig.~\ref{fig_cornerCombo_HzmSN}), remarkably locating $H_0$ in a region well in between the observations made by Planck, on one side, and SH0ES, on the other. In this case, for the deviation parameter $b$, the allowed region is such that $b = 0$ is excluded at $\sim 3\sigma$ C.L. Similar results are obtained when $H(z)$ is combined with the growth sample and Pantheon with the growth sample (Fig.~\ref{fig_cornerCombo_Hzfs8mSN}), although for the latter case, $H_0$ is not well constrained. 

As expected, the joint fit of the three data samples delivers the strongest constraints on the considered parameters (gold contours and lines in Figs.~\ref{fig_cornerCombo_Hzfs8mSN} and \ref{fig_cornerAll}, and last row of Table \ref{tab_MCMCintervals}). Up to the second order of perturbative expansion on the deviation parameter $b$, the proposed $f(R)$ model appropriately reproduces the data (Figs.~\ref{fig_hz_dataCompare} - \ref{fig_fs8z_dataCompare}) with 
\begin{align*}
    H_0         &= 71.1^{+2.5}_{-2.2} \,\,\, \rm{km/s/Mpc}, & \sigma_8 &= 0.845^{+0.030}_{-0.029}, \\
    \Omega_{m0} &= 0.243^{+0.027}_{-0.026},                 & M        &= -19.45\pm0.04, \\
    b           &= 0.649^{+0.128}_{-0.142},
\end{align*}
results that indicate that for our model 
the preferred value of $b$ turns out to be higher than initially expected, and certainly $b\neq0$ at $\gtrsim 3.5\sigma$. However, this is not entirely stunning, since this has also been obtained by different authors previously. Remarkably, the results presented in Ref.~\citep{Odintsov:2024woi} also point to a preference of some modified gravity models over $\Lambda$CDM. 

Furthermore, we also looked at the predicted evolution of some interesting cosmological parameters (Section \ref{sec_cosmo-dynamics}), noticing that the effective equation of state, the deceleration parameter and the DE density exhibit the expected behavior, slightly deviating from $\Lambda$CDM. With regard to the DE EoS, although the discrepancy is more evident, its oscillatory evolution is not unexpected (it has been observed by other authors, e.g., \citep{Granda:2020ikv,Granda:2022ttn,Zhao:2017cud,Escamilla:2024fzq}), and leads to the conclusion that additional terms in the perturbative expansion should diminish the observable difference with $\Lambda$CDM.

By performing the Om diagnostic, and using the BF values of the constrained parameters, we have observed that the proposed $f(R)$ model predicts a DE that behaves like a cosmological constant at early times $(z \gtrsim 3.5)$ and for the near future $(z < 0)$, while at current and late times, the DE exhibits a quintessence-like evolution, in agreement with the results discussed above regarding $w_{\rm{DE}}$.

Finally, as an evaluation of the statistical analysis performed in this study, and a tool to compare different models, we implemented the AIC and BIC information criteria (Section \ref{sec_IC}), the results of which are presented in Table \ref{tab_InfoCrit}. We found that, depending on the data sample analyzed, the IC can be lower or larger for $\Lambda$CDM than for the $f(R)$ model proposed here, and that the largest differences ($|\Delta IC|$) are obtained when the Pantheon sample is used individually and in a joint statistical analysis. However, as in the other cases $2 \lesssim |\Delta IC| \lesssim 6$, the results indicate that the preference over one model or the other is modest, and the two models are essentially compatible.

\section*{Acknowledgements}
A.~O.~is supported by \emph{Patrimonio Autónomo--Fondo Nacional de Financiamiento para la Ciencia, la Tecnología y la Innovación Francisco Jos\'e de Caldas (MINCIENCIAS--COLOMBIA)} Grant No.~110685269447 RC-80740-465-2020, projects 69723 and 69553. M.~A.~A.~and A.~O.~express their gratitude to Ricardo Vega (Universidad del Atl\'antico) for allowing them to use the computer resources of his Laboratory for the initial stages of the MCMC analyses.


\bibliographystyle{model1-num-names}

\bibliography{fR_biblio_2024}

\end{document}